\documentstyle[11pt]{article}
\begin{document}
\begin{titlepage}
\hfill{UQMATH-93-11}
\vskip.3in
\begin{center}
\renewcommand{\thefootnote}{\fnsymbol{footnote}}
{\huge Quantum Affine Lie Algebras, Casimir Invariants and Diagonalization
of the Braid Generator \footnote{Work supported by the
Australian Research Council.}}
\vskip.3in
{\Large Mark D. Gould} and {\Large Yao-Zhong Zhang}
\vskip.3in
{\large Department of Mathematics, University of Queensland, Brisbane,
Qld 4072, Australia}
\end{center}
\vskip.6in
\begin{center}
{\bf Abstract:}
\end{center}
Let $U_q(\hat{\cal G})$ be an infinite-dimensional quantum affine Lie algebra.
A family of central elements or Casimir invariants are constructed and their
eigenvalues computed in any integrable irreducible highest weight
representation. These eigenvalue formulae are shown to absolutely convergent
when the deformation parameter $q$ is such that $|q|>1$. It is proven
that the universal R-matrix $R$ of $U_q(\hat{\cal G})$ satisfies the
celebrated conjugation relation $R^\dagger=TR$ with $T$ the usual twist map.
As applications, the braid generator is shown to be diagonalizable
on arbitrary tensor product modules of integrable irreducible highest
weight $U_q(\hat{\cal G})$-module and a
spectral decomposition formula for the braid generator is obtained
which is the generalization of
Reshetikhin's and Gould's forms to the present affine case.
Casimir invariants acting on a specified module are also constructed and
their eigenvalues, again absolutely convergent for $|q|>1$, computed by means
of the spectral decomposition formula.

\end{titlepage}

\section{Introduction}
It is a well-known fact that for any Kac-Moody algebra, central elements
or Casimir invariants of it universal enveloping algebra play a key
role in the study of its representation theory \cite{Kac}. For quantum
Kac-Moody algebras \cite{Drinfeld}\cite{Jimbo}
which are defined as $q$-deformations of classical
(universal enveloping) Kac-Moody algebras with a symmetrizable, generalized
Cartan matrix, corresponding Casimir invariants
and their eigenvalues are expected to be of similar importance
for both formal developments of the representation
theory and practical applications.

Casimir invariants for finite-dimensional quantum simple Lie
(super)algebras $U_q({\cal G})$ have been studied
by a numbers of authors \cite{Reshetikhin}\cite{ZGB}\cite{Lee} and
general methods for constructing these invariants were proposed \cite{ZGB}
\cite{Lee}. On the other hand, braid generators for quantum simple Lie
(super)algebras are shown to be diagonalizable on any tensor product
of irreducible
highest weight (IHW) modules of the quantum algebras, in both the
multiplicity-free \cite{Reshetikhin} case and
with finite multiplicity \cite{Gould}.
Such a diagonalized form is very useful in computing the
quantum group invariants and Yang-Baxterization \cite{Zhang et al}.

In this paper we continue the developments to the
corresponding infinite-dimensional quantum affine Lie algebras
$U_q(\hat{\cal G})$. In \cite{GZ}, we addressed some of those aspcets in a
slight different formalism. However, the relevant formulae there contain
unknown quantities such as the multiplicities in the quantum
Clebsch-Gordon decomposition and therefore they are not directly applicable
to physics. Here, among others, we determine these multiplicities in
terms of the weight spectrum of some reference representation and thus
make quite explicit the expressions for the eigenvalues.

After recalling, in section 2, some fundamentals on $U_q(\hat{\cal G})$,
we in section 3 and section 4 discuss some properties of weights and
$q$-dimensions for $U_q(\hat{\cal G})$. In section 5, we examine the quantum
Clebsch-Gordon decomposition for tensor product modules
and express the multiplicities in the decomposition in terms of the
weight spectrum of a fixed (but arbitrarily chosen) reference integrable IHW
$U_q(\hat{\cal G})$-module. In section 6, we construct a
family of Casimir invariants acting on an {\em arbitrary} integrable IHW module
and compute their eigenvalues. These eigenvalue formulae are shown in
section 7 and the appendix to be absolutely convergent for $|q|>1$. In section
8 we show that the universal $R$-matrix of $U_q(\hat{\cal G})$
satisfies a conjugation relation. By means of this conjugation relation,
we show that braid generators are diagonzlizable on tensor
products of integrable IHW
$U_q(\hat{\cal G})$-modules, and
obtain a spectral decomposition formula for the braid generator which
is the generalization of Reshetikhin's and Gould's forms \cite{Reshetikhin}
\cite{Gould} to our affine
case. Using this spectral decomposition formula we construct
a family of Casimir invariants acting on some {\em specified} module
and compute their eigenvalues;
the eigenvalues are also absolutely convergent
for $|q|>1$. Section 9 is devoted to some remarks.

\section{Preliminaries}
To begin with, let $A=(a_{ij})_{0\leq i,j\leq r}$ be a symmetrizable,
generalized Cartan matrix in the sense of Kac \cite{Kac}.
Let $\hat{\cal G}$ denote the affine Lie algebra
associated with the corresponding symmetric Cartan  matrix $A_{\rm sym}$,
\begin{equation}
A_{\rm sym}=(a^{\rm sym}_{ij})=(\alpha_i,\alpha_j),~~~~i,j=0,1,...r
\end{equation}
where $\alpha_i$ are the simple roots of $\hat{\cal G}$ and
$r$ is the rank of the corresponding finite-dimensional simple Lie algebra
${\cal G}$. Then the quantum algebra $U_q(\hat{\cal G})$ is defined by
generators: $\{e_i,~f_i,~q^{h_i}~(i=0,1,...,r),~q^d\}$ and relations
\begin{eqnarray}
&&q^h.q^{h'}=q^{h+h'}~~~~(h,~ h'=h_i~ (i=0,1,...,r),~d)\nonumber\\
&&q^he_iq^{-h}=q^{(h,\alpha_i)} e_i\,,~~q^hf_iq^{-h}=q^{-(h,\alpha_i)}
  f_i\nonumber\\
&&[e_i, f_j]=\delta_{ij}\frac{q^{h_i}-q^{-h_i}}{q-q^{-1}}\nonumber\\
&&\sum^{1-a_{ij}}_{k=0}(-1)^k e_i^{(1-a_{ij}-k)}e_je_i^{(k)}
   =0~~(i\neq j)\nonumber\\
&&\sum^{1-a_{ij}}_{k=0}(-1)^k f_i^{(1-a_{ij}-k)}f_jf_i^{(k)}
   =0~~(i\neq j)\label{relations1}
\end{eqnarray}
where
\begin{equation}
e_i^{(k)}=\frac{e^k_i}{[k]_q!},~~~f^{(k)}_i=\frac{f^k_i}{[k]_q!}
\,,~~~[k]_q=\frac{q^k-q^{-k}}{q-q^{-1}}\,,~~~[k]_q!=[k]_q[k-1]_q\cdots [1]_q
\end{equation}

The Cartan subalgebra (CSA) of $\hat{\cal G}$ is generated by $\{h_i,\;i=0,1,
\cdots,r\,;\,d\}$. However, we will choose as the CSA of $\hat{\cal G}$
\begin{equation}
{\cal H}={\cal H}_0\bigoplus{\bf C}\,c\bigoplus{\bf C}\,d
\end{equation}
where $c=h_0+h_\psi$,~ $\psi$ is the highest root of ${\cal G}$ and
${\cal H}_0$ is a CSA of ${\cal G}$.

The algebra $U_q(\hat{\cal G})$ is a Hopf algebra with coproduct, counit and
antipode similar to the case of $U_q(\cal G)$:
\begin{eqnarray}
&&\Delta(q^h)=q^h\otimes q^h\,,~~~h=h_i,~d\,,~~~i=0,1,\cdots, r\nonumber\\
&&\Delta(e_i)=q^{-h_i/2}\otimes e_i+e_i\otimes q^{h_i/2}\nonumber\\
&&\Delta(f_i)=
q^{-h_i/2}\otimes f_i+f_i\otimes q^{h_i/2}\nonumber\\
&&S(a)=-q^{h_\rho}aq^{-h_\rho}\,,~~~a=e_i,f_i,h_i,d\label{coproduct1}
\end{eqnarray}
where $\rho$ is the half-sum of the positive roots of $\hat{\cal G}$.
We have omitted the formula for counit since it is not required.

Let $\Delta'$ be the opposite coproduct: $\Delta'=T\Delta$, where $T$ is
the twist map: $T(x\otimes y)=y\otimes x\,,~\forall x,y\in U_q(\hat{\cal G})$.
Then $\Delta$ and $\Delta'$ are related by the universal $R$-matrix $R$
in $U_q(\hat{\cal G})\otimes U_q(\hat{\cal G})$ satisfying, among others,
the relations
\begin{eqnarray}
&&\Delta'(x)R=R\Delta(x)\,,~~~~~\forall x\in U_q(\hat{\cal G})\nonumber\\
&&R^{-1}=(S\otimes I)R\,,~~~~~R=(S\otimes S)R\label{hopf}
\end{eqnarray}

We define a conjugate operation $\dagger$ and an anti-involution $\theta$ on
$U_q(\hat{\cal G})$ by
\begin{eqnarray}
&&d^\dagger=d\,,~~h_i^\dagger=h_i\,,~~e_i^\dagger=f_i\,,~~f_i^\dagger=e_i
 \,,~~~~i=0,1,...r \nonumber\\
&&\theta(q^h)=q^{-h}\,,~~\theta(e_i)=f_i\,,~~\theta(f_i)=
e_i\,,~~\theta(q)=q^{-1}
\end{eqnarray}
which extend uniquely to an algebra anti-automorphism and anti-involution on
all of $U_q(\hat{\cal G})$, respectively, so that $(ab)^\dagger=b^\dagger
a^\dagger\,,~~\theta(ab)=\theta(b)\theta(a)\,,~~\forall a,b\in U_q(\hat{\cal
G})$. Throughout this paper we will consider generic $q$ and use the notations
\begin{eqnarray}
&&(n)_q=\frac{1-q^n}{1-q}\,,~~[n]_q=\frac{q^n-q^{-n}}{q-q^{-1}}\,,~~
  q_\alpha=q^{(\alpha,\alpha)}\nonumber\\
&&{\rm exp}_q(x)=\sum_{n\geq 0}\frac{x^n}{(n)_q!}\,,~~(n)_q!=
  (n)_q(n-1)_q\,...\,(1)_q\nonumber\\
&&({\rm ad}_qx_\alpha)x_\beta=[x_\alpha\,,\,x_\beta]_q=x_\alpha x_\beta -
 q^{(\alpha\,,\,\beta)}x_\beta x_\alpha\;.
\end{eqnarray}

Following the usual convention, we denote the weight of
a representation by $\lambda\equiv (\lambda_0,\kappa,\tau)$, where $\lambda_0$
$\in {\cal H}_0^*\subset{\cal H}^*$ is a weight of ${\cal G}$ and
$\kappa=\lambda(c)\,,\,\tau=\lambda(d)$. The non-degenerate form $(~,~)$
on ${\cal H}^*$ is defined by \cite{GO}
\begin{equation}
(\lambda,\lambda')=(\lambda_0,\lambda'_0)+\kappa\,\tau'+\kappa'\,\tau
\,,~~~~{\rm for}~\lambda'\equiv (\lambda'_0,\kappa',\tau')~.
\end{equation}
With these notations we have
\begin{equation}
\rho=(\rho_0,0,0)+g\bar{\gamma}
\end{equation}
where $\rho_0$ is the half sum of positive roots of ${\cal G}$
,~$\bar{\gamma}=(0,1,0)$ and $2g=(\psi,\psi+2\rho_0)$.

We recall that $\lambda\in {\cal H}^*$ is called dominant integral iff
\begin{equation}
<\lambda,\alpha_i>\in {\bf Z_+}\,,~~~~0\leq i\leq r
\end{equation}
where $<\lambda,\alpha_i>\equiv 2(\lambda,\alpha_i)/(\alpha_i,\alpha_i)$\,;
$\lambda\in {\cal H}^*$ is called integral if
\begin{equation}
<\lambda,\alpha_i>\in {\bf Z}\,,~~~~0\leq i\leq r\;.
\end{equation}
Let $D$ denote the set of integral weights and $D_+$ the set of dominant
integral weights. For $\lambda\in D_+$, we call
\begin{equation}
D_q[\lambda]={\rm tr}(\pi_{\lambda}(q^{2h_\rho}))=\sum_{\mu\in\Pi(\lambda)}
m_\mu q^{(\mu,2\rho)}\label{d-q}
\end{equation}
the $q$-dimension of the integrable IHW representation
$\pi_\lambda$, where $m_\mu$ is the multiplicity of weight $\mu$ and
$\Pi(\lambda)$ denotes the weight spectrum of the integrable IHW module
$V(\lambda)\,,~\lambda\in D_+$. Explicitly the $q$-dimension takes the
form \cite{Kac}
\begin{eqnarray}
{D}_q[\lambda]&=&q^{2(\lambda,\rho)}
\prod_{\alpha\in \Phi^+}\left (\frac{1-q^{-2(\lambda+\rho,\alpha)}}
 {1-q^{-2(\rho,\alpha)}}\right )^{\rm mult\alpha}\nonumber\\
&=&q^{2g(\lambda,\bar{\gamma})}D^0_q[\lambda_0]
 \prod_{\alpha\in \Phi^+_1}\left (\frac{1-q^{-2(\lambda+\rho,\alpha)}}
 {1-q^{-2(\rho,\alpha)}}\right )^{\rm mult\alpha}\label{q-dimension2}
\end{eqnarray}
with $D^0_q[\lambda_0]$ given by
\begin{equation}
D^0_q[\lambda_0]=\prod_{\alpha\in\Phi^+_0}\frac{q^{(\lambda_0+\rho_0,\alpha)}
-q^{-(\lambda_0+\rho_0,\alpha)}}{q^{(\rho_0,\alpha)}-q^{-(\rho_0,\alpha)}}
\end{equation}
where $\Phi^+_0$ denotes the set of positive roots of ${\cal G}$,
mult$\alpha$ is the multiplicity of root $\alpha$, which is 1 when $\alpha$
is real and $r$ when $\alpha$ is imaginary, and
$\Phi^+$ is the set of all positive roots of $\hat{\cal G}$, finally $\Phi^+_1$
is the set of all positive roots excluding those of the form $(\alpha,0,0)\,,~
\alpha\in \Phi_0^+$.

With the coproduct (\ref{coproduct1}), the universal $R$-matrix has the general
form
\begin{equation}
R=\sum_sa_s\otimes b_s\label{99}
\end{equation}
where $\{a_s\,|\,s=1,2,\cdots\,\}$ and $\{b_s\,|\,s=1,2,\cdots\,\}$ are
bases of the subalgebras $U_q^\pm(\hat{\cal G})$ of $U_q(\hat{\cal G})$,
generated by $\{e_i,h_i~(i=0,1,\cdots, r);d\}$
and $\{f_i,h_i~(i=0,1,\cdots,r);d\}$, respectively. Then according to
\cite{Drinfeld}\cite{FR}, there exists a distinguished element associated
with (\ref{99}):
\begin{equation}
u=\sum_sS(b_s)a_s\label{u}\;,
\end{equation}
which has inverse
\begin{equation}
u^{-1}=\sum_sS^{-2}(b_s)a_s\label{u-1}
\end{equation}
and satisfies
\begin{eqnarray}
&&S^2(a)=uau^{-1}\,,~~~\forall a\in U_q(\hat{\cal G})\nonumber\\
&&\Delta(u)=(u\otimes u)(R^TR)^{-1}
\end{eqnarray}
where $R^T=T(R)$.  One can show that
$v=uq^{-2h_\rho}$ belongs to the center
of $U_q(\hat{\cal G})$ and satisfies
\begin{equation}
\Delta(v)=(v\otimes v) (R^TR)^{-1}\label{vv1}\;.
\end{equation}
Moreover, on an integrable irreducible representation of highest weight
$\lambda\equiv (\lambda_0,\kappa,\tau)\in D^+$,
the Casimir operator $v$ takes the eigenvalue \cite{ZG}
\begin{equation}
\chi_\lambda(v)=q^{-(\lambda,\lambda+2\rho)}\label{chi1}
\end{equation}
The following result is proven in \cite{ZGB}:
\vskip.1in
\noindent {\bf Proposition 2.1:} Let $V(\lambda)$ be an IHW
$U_q(\hat{\cal G})$-module with highest weight $\lambda\in D^+$.
If the operator $\Gamma\in U_q(\hat{\cal G})\otimes {\rm End}
V(\lambda)$ satisfies $\Delta_\lambda(a)\Gamma=\Gamma\Delta_\lambda(a)$,~~
$\forall a\in U_q(\hat{\cal G})$, where $\Delta_\lambda=(I\otimes \pi_\lambda)
\Delta$, then
\begin{equation}
C=(I\otimes {\rm tr})\{[I\otimes \pi_\lambda(q^{2h_\rho})]\Gamma\}
\end{equation}
belongs to the center of $U_q(\hat{\cal G})$, i.e. $C$ is a Casimir invariant
of $U_q(\hat{\cal G})$.

\section{Linkage and Regular Weights}
For generic $q$, the representation theory of
$U_q(\hat{\cal G})$ bears much similarity to
that of $\hat{\cal G}$ \cite{Rosso}\cite{ZG}. In particular, classical
and corresponding quantum representations have the same $q$-dimension and
weight spectrum. Therefore the analysis in this section are the same for both
the calssical and quantum cases.

Let ${\cal W}_0$ be the Weyl group of ${\cal G}$ and ${\cal W}$
the Weyl group of
$\hat{\cal G}$. Recall that
\begin{equation}
{\cal W}={\cal W}_0\otimes_{\rm s}{\cal T}
\end{equation}
where ${\cal T}$ is the abelian group of translations. Let
\begin{equation}
\tilde{\cal W}=\tilde{\cal W}_0\otimes_{\rm s}\tilde{{\cal T}}
\end{equation}
denote the translated Weyl group: for $w\in {\cal W}$ we define
$\tilde{w}\in\tilde{{\cal W}}$ by
\begin{equation}
\tilde{w}(\lambda)=w(\lambda+\rho)-\rho\;.
\end{equation}
Then $\tilde{\cal W}$ forms a group isomorphic to ${\cal W}$
(it is in fact conjugate to
${\cal W}$ in the group of invertible affine transformations of ${\cal H}^*$).
\vskip.1in
\noindent {\bf Definition 3.1:} We say that $\lambda\,,~\mu\in{\cal H}^*$
are linked if $\exists\tilde{w}\in\tilde{\cal W}$ such that
\begin{equation}
\tilde{w}(\lambda)=\mu\;.
\end{equation}
Equivalently, $\lambda$ and $\mu$ are linked iff $\exists w\in {\cal W}$ such
that
\begin{equation}
w(\lambda+\rho)=\mu+\rho\;.
\end{equation}
\vskip.1in
We now investigate conditions under which a given integral weight $\mu$ is
linked to a dominant integral weight. Now $\mu+\rho$ is integral, then
according to Kac \cite{Kac}, $\mu+\rho$ is conjugate under ${\cal W}$ to a
(unique) dominant integral weight $\lambda\in D_+$.
The following conditions are equivalent:
\begin{eqnarray}
{\rm (i)}&~& (\mu+\rho,\alpha)\neq 0\,,~~~~
       \forall \alpha\in \Phi_{\rm re}^+\nonumber\\
{\rm (ii)}&~&(\mu+\rho,\alpha)\neq 0\,,~~~~
       \forall \alpha\in \Phi_{\rm re}\nonumber\\
{\rm (iii)}&~&(\lambda,\alpha)\neq 0\,,~~~~
       \forall \alpha\in \Phi_{\rm re}\nonumber\\
{\rm (iv)}&~&<\lambda,\alpha>~> 0\,,~~~~
       \forall \alpha\in \Phi_{\rm re}^+\label{equiv}
\end{eqnarray}
where $\Phi_{\rm re}^+$ denotes the set of positive real roots of
$\hat{\cal G}$ and $\Phi_{\rm re}$ the set of its real roots.
The above follows from  the fact that $\Phi_{\rm re}$ is invariant
under the action of ${\cal W}$.
\vskip.1in
\noindent {\bf Definition 3.2:} We call $\mu\in{\cal H}^*$ regular provided
\begin{equation}
(\mu+\rho,\alpha)\neq 0\,,~~~~\forall\alpha\in \Phi_{\rm re}^+
\end{equation}
Otherwise $\mu$ is called irregular.
\vskip.1in
It follows from the equivalences above that if $\mu\in D$ is regular then
$<\lambda,\alpha_i>~>0\,,~~0\leq i\leq r$. Since $<\rho,\alpha_i>=1$, one
has $<\lambda-\rho,\alpha_i>\geq 0\,,~~0\leq i\leq r$. Thus if $\mu$ is
regular, then $\lambda-\rho\in D_+$ and hence $\mu+\rho$ is
${\cal W}$-conjugate
to $\lambda=(\lambda-\rho)+\rho$ or $\mu$ is $\tilde{\cal W}$-conjugate to
$\lambda-\rho\in D_+$.

Thus $\mu\in D$ regular $\Rightarrow \mu$ linked to a dominant weight.
Coversely, if $\mu$ is linked to a dominant weight $\lambda\in D_+$,
i.e. $w(\mu+\rho)=\lambda+\rho$, then $\mu$ is regular. This follows
since $<\lambda+\rho,\alpha>~>0\,,~~\forall\alpha\in\Phi_{\rm re}^+$ and
$w$ permutes the real roots. We thus have the following
\vskip.1in
\noindent {\bf Proposition 3.1:} An integral weight $\mu$ is
$\tilde{\cal W}$-conjugate to a dominant integral weight iff $\mu$ is regular.
\vskip.1in
As we will see, this result is important for simplifying the eigenvalue
formulae for Casimir invariants of $U_q(\hat{\cal G})$.
We shall also need some results on the
transformation properties of the $q$-dimension $D_q[\lambda]$ under
$\tilde{\cal W}$.

\section{Transformation Properties of $q$-Dimension}
Our aim here is to investigate the transformation properties of the
$q$-dimension function under $\tilde{\cal W}$. First of all, for elements
$\tilde{w}_0\in\tilde{\cal W}_0$, we have
\begin{eqnarray}
\tilde{w}_0D_q[\lambda]&=&D_q[\tilde{w}_0^{-1}(\lambda)]\nonumber\\
 &\stackrel{(\ref{q-dimension2})}{=}&q^{2g(\lambda,\bar{\gamma})}D^0_q
  [\tilde{w}_0^{-1}(\lambda_0)]
\prod_{\alpha\in \Phi^+_1}\left (\frac{1-q^{-2(\lambda+\rho,\alpha)}}
 {1-q^{-2(\rho,\alpha)}}\right )^{\rm mult\alpha}
\end{eqnarray}
where we have used the fact that $\bar{\gamma}$ is orthognal to $\rho$ and is
invariant under ${\cal W}_0$ together with the fact that
\begin{equation}
(\tilde{w}_0^{-1}(\lambda)+\rho,\alpha)=(w_0^{-1}(\lambda+\rho),
\alpha)=(\lambda+\rho,w_0(\alpha))
\end{equation}
and the fact that $\Phi_1^+$ is permuted by the action of $w_0\in {\cal W}_0$.
Finally,
\begin{eqnarray}
D^0_q[\tilde{w}_0^{-1}(\lambda_0)]&=&
  \prod_{\alpha\in\Phi^+_0}\frac{q^{({w}_0^{-1}(\lambda_0+\rho_0),\alpha)}
  -q^{-({w}_0^{-1}(\lambda_0+\rho_0),\alpha)}}
  {q^{(\rho_0,\alpha)}-q^{-(\rho_0,\alpha)}}\nonumber\\
&=&\prod_{\alpha\in\Phi^+_0}\frac{q^{(\lambda_0+\rho_0,w_0(\alpha))}
  -q^{-(\lambda_0+\rho_0,w_0(\alpha))}}
  {q^{(\rho_0,\alpha)}-q^{-(\rho_0,\alpha)}}\nonumber\\
&=&(-1)^{l(w_0)}\;\prod_{\alpha\in\Phi^+_0}\frac{q^{(\lambda_0+\rho_0,\alpha)}
  -q^{-(\lambda_0+\rho_0,\alpha)}}
  {q^{(\rho_0,\alpha)}-q^{-(\rho_0,\alpha)}}\nonumber\\
&=&{\rm sign}w_0\;D_q^0[\lambda_0]
\end{eqnarray}
where $l(w_0)$ is the number of positive roots of ${\cal G}$ sent
to negative roots by $w_0$ and ${\rm sign}w_0=(-1)^{l(w_0)}$ is the sign of
the Weyl group element $w_0\in {\cal W}_0$.

Combining the above results we arrive at the transformation law:
\vskip.1in
\noindent {\bf Proposition 4.1:}
\begin{equation}
\tilde{w}_0D_q[\lambda]\equiv D_q[\tilde{w}_0^{-1}(\lambda)]
  ={\rm sign}w_0\;D_q[\lambda]\,,~~~~\forall w_0\in {\cal W}_0\;.
\end{equation}

In order to examine the full transformation properties of the $q$-dimension,
we first recall the Kac-Weyl character formula. The $q$-character
of an integrable IHW module $V(\lambda)\,,~\lambda\in D_+$ is defined by
\begin{equation}
{\rm ch}_q(\lambda)=\sum_{\mu\in\Pi(\lambda)}m_\mu\,q^\mu\label{character}
\end{equation}
where $m_\mu$ is the multiplicity of weight $\mu$ in $V(\lambda)$. From
(\ref{character}) one deduces the Kac-Weyl character formula \cite{Kac},
\begin{equation}
{\rm ch}_q(\lambda)=P_q^{-1}\sum_{w\in {\cal W}}
 {\rm sign}w\,q^{w(\lambda+\rho)-\rho}\label{kac-weyl-character}
\end{equation}
where ${\rm sign}w$ is the sign of the Weyl group element $w\in{\cal W}$ and
$P_q$ is the Kac-Weyl denominator function,
\begin{equation}
P_q=\prod_{\alpha\in\Phi^+}\left (1-q^{-\alpha}\right )^{{\rm mult}\alpha}\;.
\end{equation}
Applying (\ref{kac-weyl-character}) to the trivial irrep with the highest
weight $\lambda=(0,0,0)$ gives rise to the Kac-Weyl denominator formula,
\begin{equation}
P_q=\sum_{w\in {\cal W}}{\rm sign}w\,q^{w(\rho)-\rho}\;.
\end{equation}

It is easily seen from (\ref{character}) that the $q$-dimension (\ref{d-q})
may expressed as
\begin{equation}
D_q[\lambda]=f_{2\rho}\cdot {\rm ch}_q(\lambda)
\end{equation}
where $f_\mu$ is defined as
\begin{equation}
f_\mu\cdot q^\nu=q^{(\mu,\nu)}\;.
\end{equation}
We now give an alternative but equivalent formula for the $q$-dimension,
\begin{equation}
D_q[\lambda]=\frac{q^{-2(\rho,\rho)}}{P_q(2\rho)}\sum_{t\in {\cal T}}
  q^{2(t(\lambda+\rho),\rho)}\prod_{\alpha\in \Phi_0^+}\left (
  1-q^{-2(t(\lambda+\rho),\alpha)}\right )\label{q-dimension3}
\end{equation}
where
\begin{equation}
P_q(2\rho)=f_{2\rho}\cdot P_q=\prod_{\alpha\in\Phi^+}\left (
 1-q^{-(\alpha,2\rho)}\right )^{{\rm mult}\alpha}
\end{equation}
\vskip.1in
\noindent {\bf Proof of (\ref{q-dimension3}):}
\begin{eqnarray}
D_q[\lambda]&=&f_{2\rho}\cdot {\rm ch}_q(\lambda)
  =f_{2\rho}\cdot\frac{1}{P_q}\sum_{w\in {\cal W}}
  {\rm sign}w\,q^{w(\lambda+\rho)-\rho}\nonumber\\
&=&\frac{q^{-2(\rho,\rho)}}{P_q(2\rho)}\sum_{w\in {\cal W}}{\rm sign}w\,
  q^{(w(\lambda+\rho),2\rho)}\nonumber\\
&=&\frac{q^{-2(\rho,\rho)}}{P_q(2\rho)}\sum_{t\in {\cal T}}
  \sum_{w_0\in {\cal W}_0}
  {\rm sign}w_0\,q^{(w_0t(\lambda+\rho),2\rho)}\nonumber\\
&=&\frac{q^{-2(\rho,\rho)}}{P_q(2\rho)}\sum_{t\in {\cal T}}
  \sum_{w_0\in {\cal W}_0}{\rm sign}w_0^{-1}
  \,q^{(w_0^{-1}t(\lambda+\rho),2\rho)}\nonumber\\
&=&\frac{q^{-2(\rho,\rho)}}{P_q(2\rho)}\sum_{t\in {\cal T}}
  \sum_{w_0\in {\cal W}_0}
  {\rm sign}w_0\,q^{(w_0^{-1}t(\lambda+\rho),2\rho)}\nonumber\\
&=&\frac{q^{-2(\rho,\rho)}}{P_q(2\rho)}\sum_{t\in {\cal T}}
  \sum_{w_0\in {\cal W}_0}
  {\rm sign}w_0\,q^{(2t(\lambda+\rho),w_0(\rho))}\nonumber\\
&=&\frac{q^{-2(\rho,\rho)}}{P_q(2\rho)}\sum_{t\in {\cal T}}f_{2t(\lambda+\rho)}
  \cdot\sum_{w_0\in {\cal W}_0}{\rm sign}w_0
  \,q^{w_0(\rho)}\nonumber\\
&=&\frac{q^{-2(\rho,\rho)}}{P_q(2\rho)}\sum_{t\in {\cal T}}f_{2t(\lambda+\rho)}
  \cdot q^{g\gamma+\rho_0}\sum_{w_0\in {\cal W}_0}{\rm sign}w_0
  \,q^{w_0(\rho_0)-\rho_0}\nonumber\\
&=&\frac{q^{-2(\rho,\rho)}}{P_q(2\rho)}\sum_{t\in {\cal T}}f_{2t(\lambda+\rho)}
  \cdot q^\rho\prod_{\alpha\in\Phi_0^+}\left (1-q^{-\alpha}\right )
\end{eqnarray}
where use has been made of the fact that ${\rm sign}w_0^{-1}={\rm sign}w_0$.
We immediately see that (\ref{q-dimension3}) is satisfied.~~~~$\Box$

We are now in the position to investigate the transformation properties of
the $q$-dimension under $\tilde{\cal T}$. We have
\vskip.1in
\noindent {\bf Proposition 4.2:} For $t_1\in {\cal T}$,
\begin{equation}
\tilde{t}_1D_q[\lambda]=D_q[\lambda]\;.
\end{equation}
\vskip.1in
\noindent{\bf Proof:}
\begin{eqnarray}
\tilde{t}_1D_q[\lambda]&=&D_q[\tilde{t}^{-1}_1(\lambda)]
   =D_q[t^{-1}_1(\lambda+\rho)-\rho]\nonumber\\
&=&\frac{q^{-2(\rho,\rho)}}{P_q(2\rho)}\sum_{t\in {\cal T}}
  q^{2(tt^{-1}_1(\lambda+\rho),\rho)}\prod_{\alpha\in \Phi_0^+}\left (
  1-q^{-2(tt^{-1}_1(\lambda+\rho),\alpha)}\right )\nonumber\\
&=&\frac{q^{-2(\rho,\rho)}}{P_q(2\rho)}\sum_{t\in {\cal T}}
  q^{2(t(\lambda+\rho),\rho)}\prod_{\alpha\in \Phi_0^+}\left (
  1-q^{-2(t(\lambda+\rho),\alpha)}\right )\nonumber\\
&=&D_q[\lambda]
\end{eqnarray}
where use has been made of the fact that $tt_1^{-1}$ must run through the
elements of ${\cal T}$ as $t$ runs through ${\cal T}$.~~~~~$\Box$

We may now determine the transformation
properties of the $q$-dimension under the full translated Weyl group
$\tilde{\cal W}$.
\vskip.1in
\noindent{\bf Proposition 4.3:}
For $w=w_0t\,,~w_0\in {\cal W}_0\,,~t\in {\cal T}$,
\begin{equation}
\tilde{w}D_q[\lambda]\equiv D_q[\tilde{w}^{-1}(\lambda)]
  ={\rm sign}w\,D_q[\lambda]
\end{equation}
\vskip.1in
\noindent {\bf Proof:} it follows from propositions 4.1 and 4.2 as well as
the fact that ${\rm sign}t=1$ and ${\rm sign}w={\rm sign}w_0$.

We will apply this transformation property below in deriving a suitable
formula for the Casimir invariants of $U_q(\hat{\cal G})$.

\section{Clebsch-Gordon Decomposition}
Let $V(\Lambda)$ be a fixed (but arbitrarily chosen)
reference $U_q({\cal G})$-module with $\Lambda
\in D^+$ and we consider the reduction
of $V(\mu)\otimes V(\Lambda)$ with $V(\mu)\,,~\mu\in D^+$ being an
arbitrary IHW $U_q(\hat{\cal G})$-module.
We have shown in \cite{ZG} that the tensor product $V(\lambda)\otimes
V(\nu)$ of the integrable IHW $U_q(\hat{\cal G})$-modules
$V(\lambda)$ and $V(\nu)$ is completely reducible and the
irreducible components are integrable highest weight representations. This
means
that we may decompose the tensor product $V(\mu)\otimes V(\Lambda)$
according to
\begin{equation}
V(\mu)\otimes V(\Lambda)=\bigoplus_{
\hat{\nu}\in D^+}m_{\hat{\nu}}V(\hat{\nu})
\end{equation}
where $m_{\hat{\nu}}$ are the multiplicities of the modules
$V(\hat{\nu})$ in the above decomposition.

Here we determine the multiplicity $m_{\hat{\nu}}$
in terms of the weight spectrum of
the reference representation $\pi_\Lambda$. To this end, we introduce the
set of weights $\Pi_\mu(\Lambda)\subseteq \Pi(\Lambda)$ defined by
\begin{equation}
\Pi_\mu(\Lambda)=\{\nu\in\Pi(\Lambda)|\mu+\nu~ {\rm is~regular}\}\;.
\end{equation}
Then for $\nu\in\Pi_\mu(\Lambda)$, there exists $w_\nu\in {\cal W}$ such that
$\tilde{w}_\nu(\mu+\nu)$ is dominant; we write $\tilde{w}_\nu(\mu+\nu)
=\hat{\nu}\in D^+$, i.e. $\hat{\nu}$ and $\mu+\nu$ are linked. We then
write $\mu+\nu\sim\hat{\nu}$. We note that $\Pi(\Lambda)$ may be
partitioned
\begin{equation}
\Pi(\Lambda)=\Pi_\mu(\Lambda)\cup \Pi^\perp_\mu(\Lambda)
\end{equation}
where
\begin{eqnarray}
\Pi^\perp_\mu(\Lambda)&=&\{\nu\in\Pi(\Lambda)|\mu+\nu~
  {\rm is~irregular}\}\nonumber\\
&=&\{\nu\in\Pi(\Lambda)|(\mu+\nu+\rho,\alpha)=0\,~~
  {\rm for~some}~\alpha\in\Phi_{\rm re}\}\;.
\end{eqnarray}
We now state
\vskip.1in
\noindent {\bf Proposition 5.1:}
\begin{equation}
m_{\hat{\nu}}=\sum_{\stackrel{\nu\in\Pi(\Lambda)}{\hat{\nu}\sim\mu+\nu}}
m_\nu\,{\rm sign}w_\nu
\end{equation}
where $w_\nu$ is such that $w_\nu(\mu+\nu+\rho)=\hat{\nu}+\rho$.
\vskip.1in
\noindent {\bf Proof:} We compute the quantity ${\rm ch}_q(\mu)\cdot
{\rm ch}_q(\Lambda)$:
\begin{eqnarray}
{\rm ch}_q(\mu)\cdot{\rm ch}_q(\Lambda)&=&\sum_{\nu\in \Pi(\Lambda)}
  m_\nu\,q^\nu\cdot P_q^{-1}\sum_{w\in {\cal W}}{\rm sign}w\,q^{w(\mu+\rho)
  -\rho}\nonumber\\
&=&P_q^{-1}\cdot \sum_{\nu\in \Pi(\Lambda)}
  m_\nu\,\sum_{w\in {\cal W}}{\rm sign}w\,q^{\nu+w(\mu+\rho)-\rho}\nonumber\\
&=&P_q^{-1}\cdot \sum_{\nu\in \Pi(\Lambda)}
  m_\nu\,\sum_{w\in {\cal W}}{\rm sign}w\,q^{w(\mu+\nu+\rho)-\rho}
\end{eqnarray}
where we have replaced $\nu$ by $w(\nu)$ and used the fact that
$m_\nu=m_{w(\nu)}$, i.e. ${\cal W}$-conjugate weights occur with the same
multiplicity in an irrep. $V(\Lambda)$.

Using the above formula we immediately obtain
\begin{equation}
{\rm ch}_q(\mu)\cdot{\rm ch}_q(\Lambda)=\sum_{\hat{\nu}\in D^+}
m_{\hat{\nu}}\,{\rm ch}_q(\hat{\nu})
\end{equation}
with the multiplicity $m_{\hat{\nu}}$ given by proposition 5.1.~~~~$\Box$
\vskip.1in
\noindent{\bf Remarks:} (i) Proposition (5.1) is analogous to the
Klimyk-Racah formula for finite dimensional simple Lie algebras.\\
(ii) Closer inspection of the above computations shows that the multiplicity
$m_{\hat{\nu}}$ is zero unless $\hat{\nu}=\mu+\lambda$ for some weight
$\lambda\in \Pi(\Lambda)$.

\section{Casimir Invariants and Eigenvalue Formulae}
We construct a family of Casimir invariants for $U_q(\hat{\cal G})$.
We state our result in the following
\vskip.1in
\noindent{\bf Proposition 6.1:} Let $\Gamma$ be the operator
\begin{equation}
\Gamma=(I\otimes\pi_{\Lambda})R^TR\label{gamma}\;.
\end{equation}
Then the operators $C^{\Lambda}_m$ defined by
\begin{equation}
C^{\Lambda}_m=(I\otimes{\rm tr})\{[I\otimes\pi_{\Lambda}(q^{2h_\rho})]
\Gamma^m\}\,,~~~~~m\in{\bf Z}^+\label{propos.2}
\end{equation}
are Casimir invariants of $U_q(\hat{\cal G})$. Acting on an
integrable IHW $U_q(\hat{\cal G})$-module
$V(\mu)$ with highest weight $\mu$, the
$C_m^{\Lambda}$ take the following eigenvalues
\begin{equation}
\chi_\mu(C^{\Lambda}_m)=\sum_{\nu\in\Pi(\Lambda)}m_\nu\,q^{-m\beta_\nu(\mu)}
\cdot \frac{D_q[\mu+\nu]}{D_q[\mu]}
\,,~~~~~~m\in{\bf Z}^+\label{chi-mu}
\end{equation}
with
\begin{equation}
\beta_\nu(\mu)=(\Lambda,\Lambda+2\rho)-(\nu,\nu+2(\mu+\rho))\label{proof1}\;.
\end{equation}
\vskip.1in
\noindent {\bf Remark 6.1:} Unlike the relevant formula in \cite{GZ} which
is in terms of the unknown multiplicities and is thus nonexplicit,
the formula (\ref{chi-mu})
determines a function on ${\cal H}^*$,
well defined on all regular weights.
It requires only a knowledge of the weight spectrum of the reference irrep.
$\pi_\Lambda$, regardless of $\pi_\mu$ and therefore is a very explicit
formula.
\vskip.1in
\noindent{\bf Proof:} The statement that $C^{\Lambda}_m$ are Casimir
invariants is easy to see: since $\Gamma$ satisfies $\Delta_{\Lambda}(a)
\Gamma=\Gamma\Delta_{\Lambda}(a)\,~\forall a\in U_q(\hat{\cal G})$, so do its
higher powers; thus by proposition 2.1, $C^{\Lambda}_m$ must be Casimir
inveriants of $U_q(\hat{\cal G})$. We now come to the second part of the
proposition. By (\ref{vv1}) we have
\begin{equation}
\Gamma=(I\otimes\pi_{\Lambda})R^TR=(I\otimes \pi_{\Lambda})
((v\otimes v)\Delta(v^{-1}))=(v\otimes\pi_{\Lambda}(v))\partial
(v^{-1})\label{2}
\end{equation}
where $\partial$ is the algebra homomorphism defined by
\begin{eqnarray}
&&\partial \,:\, U_q(\hat{\cal G})\longrightarrow U_q(\hat{\cal G})\otimes
  {\rm End}V(\Lambda)\nonumber\\
&&\partial (v^{-1})=(I\otimes\pi_{\Lambda})\Delta (v^{-1})\;.
\end{eqnarray}

Now it follows from (\ref{chi1}) that on irrep
$V(\hat{\nu})\subseteq V(\mu)
\otimes V(\Lambda)$\,,~$\Gamma$ in (\ref{2}) takes the value:
\begin{equation}
\chi_{\hat{\nu}}(\Gamma)=\chi_\mu(v)\chi_\Lambda(v)\chi_{\hat{\nu}}(v^{-1})
=q^{-\alpha_{\hat{\nu}}(\mu)}
\end{equation}
with
\begin{equation}
\alpha_{\hat{\nu}}(\mu)=(\Lambda,\Lambda+2\rho)+(\mu,\mu+2\rho)-
   (\hat{\nu},\hat{\nu}+2\rho)\;.
\end{equation}
Let $P[\hat{\nu}]$ be the central projections:
\begin{equation}
P[\hat{\nu}] (V(\mu)\otimes V(\Lambda))=V(\hat{\nu})\;.
\end{equation}
With the help of the projection operators, $\Gamma^m$ can be expressed as
\begin{equation}
\Gamma^m=\sum_{\hat{\nu}\in D^+}q^{-m\alpha_{\hat{\nu}}(\mu)}P[\hat{\nu}]\;.
\end{equation}
Inserting them into (\ref{propos.2}) and noting that $C_m^{\Lambda}$ are
Casimir invariants we find
\begin{equation}
\chi_\mu (C_m^{\Lambda})=\sum_{\hat{\nu}\in D^+}q^{-m\alpha_{\hat{\nu}}(\mu)}
(I\otimes{\rm tr})\{(I\otimes\pi_{\Lambda}(q^{2h_\rho}))P[\hat{\nu}]\}
\end{equation}
which gives, after some effort
\begin{eqnarray}
\chi_\mu(C_m^{\Lambda})
&=&\sum_{\hat{\nu}\in D^+}m_{\hat{\nu}}\,q^{-m\alpha_{\hat{\nu}}(\mu)}
\frac{D_q[\hat{\nu}]}{D_q[\mu]}\nonumber\\
&=&\sum_{\hat{\nu}\in D^+}\sum_{
\stackrel{\nu\in\Pi(\Lambda)}
{\mu+\nu\sim\hat{\nu}}
}m_\nu\,{\rm sign}w_\nu\,q^{-m\alpha_{\hat{\nu}}(\mu)}
\frac{D_q[\hat{\nu}]}{D_q[\mu]}
\end{eqnarray}
where, as before, $w_\nu\in{\cal W}$ is such that $w_\nu(\mu+\nu+\rho)
=\hat{\nu}+\rho$.

It is easy to see that for $\mu+\nu\sim\hat{\nu}$ we have
\begin{equation}
\alpha_{\hat{\nu}}(\mu)\equiv\beta_\nu(\mu)=
(\Lambda,\Lambda+2\rho)-(\nu,\nu+2(\mu+\rho))\label{beta}\;.
\end{equation}
Moreover, since $\tilde{w}_\nu(\mu+\nu)=\hat{\nu}$, we have, from proposition
4.3,
\begin{equation}
D_q[\mu+\nu]=D_q[\tilde{w}^{-1}_\nu(\hat{\nu})]={\rm sign}w^{-1}_\nu
\,D_q[\hat{\nu}]={\rm sign}w_\nu\,D_q[\hat{\nu}]\;.
\end{equation}
We therefore arrive at
\begin{eqnarray}
\chi_\mu(C^\Lambda_m)&=&\sum_{\hat{\nu}\in D^+}\sum_{
\stackrel{\nu\in\Pi(\Lambda)}
{\mu+\nu\sim\hat{\nu}}
  }m_\nu\,q^{-m\beta_\nu(\mu)}
  \frac{D_q[\mu+\nu]}{D_q[\mu]}\nonumber\\
&=&\sum_{\nu\in \Pi_\mu(\Lambda)}
  m_\nu\,q^{-m\beta_\nu(\mu)}
  \frac{D_q[\mu+\nu]}{D_q[\mu]}
\end{eqnarray}
The sum above is over all weights $\nu\in\Pi(\Lambda)$ such that $\mu+\nu$
is regular. However, for irregular $\mu+\nu$, one has $D_q[\mu+\nu]\equiv 0$.
Therefore, the sum above may be extended over all weights $\nu$ in
$\Pi(\Lambda)$ to give the eigenvalue formula (\ref{chi-mu}).~~~~~$\Box$

On the other hand, for the first order Casimir operator $C^{\Lambda}_1$
\begin{equation}
C^{\Lambda}_1=(I\otimes{\rm tr})\{[I\otimes\pi_{\Lambda}(q^{2h_\rho})]
\Gamma\}=\sum_{s,s'}{\rm tr}(\pi_{\Lambda}(q^{2h_\rho}a_sb_{s'}))b_sa_{s'}
\end{equation}
its eigenvalue on
integrable IHW $U_q(\hat{\cal G})$-module $V(\mu)$ with
highest weight $\mu$ can be computed in the direct way \cite{GZ}, giving
\begin{equation}
\chi_\mu(C^{\Lambda}_1)=\sum_{\nu\in \Pi(\Lambda)}\,
m_\nu\;q^{2(\nu,\mu+\rho)}\label{eigenvalue1}\;.
\end{equation}
By comparing (\ref{chi-mu}) with (\ref{eigenvalue1}), we arrive at
the interesting identity
\begin{equation}
q^{(\Lambda,\Lambda+2\rho)}\sum_{\nu\in\Pi(\Lambda)}m_\nu\,q^{2(\nu,\mu+\rho)}
=\sum_{\nu\in\Pi(\Lambda)}m_\nu\,q^{(\nu,\nu+2(\mu+\rho))}
\frac{D_q[\mu+\nu]}{D_q[\mu]}\;.
\end{equation}

\section{String Functions and Convergence}
We introduce the set of $\delta$-maxmial weights $M(\Lambda)$:
\begin{equation}
M(\Lambda)=\{\nu\in\Pi(\Lambda)|\nu+\delta\not\in\Pi(\Lambda)\}
\end{equation}
where $\delta=(0,0,1)$ is the minimal imaginary positive root. Then the
weight spectrum of $\Pi(\Lambda)$ is given by
\begin{equation}
\Pi(\Lambda)=\cup_{\nu\in\Pi(\Lambda)}\{\nu-s\delta\,|\,s\in{\bf Z}_+\}\;.
\end{equation}
We note that for any weight $\lambda$
\begin{equation}
D_q[\lambda-s\delta]=q^{-2s(\delta,\rho)}D_q[\lambda]=q^{-2sg}
D_q[\lambda]\label{proof-d}\;.
\end{equation}
Moreover,
\begin{equation}
\beta_{\nu-s\delta}(\mu)=\beta_\nu(\mu)+2s(\delta,\mu+\nu+\rho)\stackrel{
(\delta,\delta)=0}{=}\beta_\nu(\mu)+2s(\delta,\mu+\Lambda+\rho)\label{proof}\;.
\end{equation}
Thanks to these two relations (\ref{chi-mu}) becomes,
\begin{equation}
\chi_\mu(C^\Lambda_m)=\sum_{\nu\in M(\Lambda)}q^{-m\beta_\nu(\mu)}
\frac{D_q[\mu+\nu]}{D_q[\mu]}\,\sum_{s=0}^\infty m_{\nu-s\delta}\,
q^{-2s\{g+m(\Lambda+\mu+\rho,\delta)\}}\label{chi-s}\;.
\end{equation}
In terms of string functions \cite{Kac}
\begin{equation}
S^\Lambda_\nu(q)=\sum_{s=0}^\infty m_{\nu-s\delta}\,q^{-s}\,,~~~\nu\in
M(\Lambda)
\end{equation}
which are absolutely covergent for $|q|>1$, eq.(\ref{chi-s}) reads
\begin{equation}
\chi_\mu(C^\Lambda_m)=\sum_{\nu\in M(\Lambda)}
S^\Lambda_\nu(q^{2\{g+m(g+\kappa_\Lambda+\kappa_\mu)\}})\,q^{-m\beta_\nu(\mu)}
\frac{D_q[\mu+\nu]}{D_q[\mu]}\label{chi-mu-string}
\end{equation}
where use has been made of the notations:
$\Lambda\equiv (\Lambda_0,\kappa_\Lambda,
\tau_\Lambda)$,~$\kappa_\Lambda\equiv \Lambda(c)\,,~\tau_\Lambda\equiv
\Lambda(d)$ and $\mu\equiv (\mu_0,\kappa_\mu,\tau_\mu)$,~$\kappa_\mu\equiv
\mu(c)\,,~\tau_\mu\equiv\mu(d)$.

The eigenvalue formulae (\ref{chi-mu}) are invariant under $\tilde{\cal W}$,
\begin{equation}
\chi_{\tilde{w}(\mu)}(C^\Lambda_m)=\chi_\mu(C^\Lambda_m)\,,~~~
\tilde{w}\in\tilde{\cal W}\label{chi-w1}
\end{equation}
To see this, we compute
\begin{eqnarray}
\chi_{\tilde{w}(\mu)}(C^\Lambda_m)&=&\sum_{\nu\in M(\Lambda)}
  S^\Lambda_\nu(q^{2\{g+m(g+\kappa_\Lambda+\kappa_{\tilde{w}(\mu))}\}})\,
  q^{-m\beta_\nu(\tilde{w}(\mu))}
  \frac{D_q[\tilde{w}(\mu)+\nu]}{D_q[\tilde{w}(\mu)]}\nonumber\\
&=&\sum_{\nu\in M(\Lambda)}
  S^\Lambda_\nu(q^{2\{g+m(g+\kappa_\Lambda+\kappa_\mu)\}})\,
  q^{-m\beta_\nu(\tilde{w}(\mu))}
  \frac{D_q[\tilde{w}(\mu)+\nu]}{D_q[\tilde{w}(\mu)]}\label{chi-w2}
\end{eqnarray}
where we have used
\begin{equation}
\kappa_{\tilde{w}(\mu)}\equiv\tilde{w}(\mu)(c)\equiv(\tilde{w}(\mu),\delta)
=(w(\mu+\rho)-\rho,\delta)=\kappa_\mu
\end{equation}
due to the invariance of $\delta$ under ${\cal W}$. It is easily established
from (\ref{beta}) that
\begin{equation}
\beta_\nu(\tilde{w}(\mu))=\tilde{w}^{-1}\cdot\beta_\nu(\mu)=
\beta_{w^{-1}(\nu)}(\mu)\;.
\end{equation}
Moreover from proposition 4.3, one has
\begin{eqnarray}
&&D_q[\tilde{w}(\mu)]=\tilde{w}^{-1}\cdot D_q[\mu]={\rm sign}w^{-1}D_q[\mu]
 ={\rm sign}w\,D_q[\mu]\nonumber\\
&&D_q[\tilde{w}(\mu)+\nu]=D_q[w(\mu+\rho)-\rho+\nu]=D_q[w(\mu+w^{-1}(\nu)
 +\rho)-\rho]\nonumber\\
&&~~~~~~~~~~=D_q[\tilde{w}(\mu+w^{-1}(\nu))]={\rm sign}w\,D_q
 [\mu+w^{-1}(\nu)]\;.
\end{eqnarray}
Substituting these into the r.h.s. of (\ref{chi-w2}) leads to
\begin{equation}
\chi_{\tilde{w}(\mu)}(C^\Lambda_m)=\sum_{\nu\in M(\Lambda)}
  S^\Lambda_\nu(q^{2\{g+m(g+\kappa_\Lambda+\kappa_\mu)\}})\,
  q^{-m\beta_{w^{-1}(\nu)}(\mu)}
  \frac{D_q[\mu+w^{-1}(\nu)]}{D_q[\mu]}\label{chi-w3}\;.
\end{equation}
With the help of the Weyl group invariance of the string functions \cite{Kac},
\begin{equation}
S^\Lambda_\nu(q)=S^\Lambda_{w(\nu)}(q)\,~~~\forall w\in{\cal W}
\end{equation}
we immediately see that the r.h.s. of (\ref{chi-w3}) equals to $\chi_\mu(
C^\Lambda_m)$. Thus we have established the symmetry (\ref{chi-w1}) which
implies that the eigenvalues determine $\tilde{\cal W}$-invariant
functions on (the regular elements of) ${\cal H}^*$ in direct analogy with
the case for finite-dimensional simple Lie algebras.

Now note that the $q$-dimension (\ref{q-dimension2}) may written as
\begin{eqnarray}
&&D_q[(\lambda_0,\kappa,\tau)]=q^{2g\tau}\,\bar{D}_q[(\lambda_0,\kappa,0)]
 \,, \nonumber\\
&&\bar{D}_q[(\lambda_0,\kappa,0)]=D^0_q[\lambda_0]\prod_{n=1}^\infty
\left (\frac{1-q^{-2n(\kappa+g)}}{1-q^{-2ng}}\right )^r
\prod_{\alpha\in
\Phi_0}\prod_{n=1}^\infty\frac{1-q^{-2(\lambda_0+\rho_0,\alpha)
-2n(\kappa+g)}}{1-q^{-2(\rho_0,\alpha)-2ng}}\label{q-dimension1}
\end{eqnarray}
which is absolutely convergent for $|q|\,>\,1$.
It follows (see Appendix for proof) that the eigenvalue
formulae (\ref{chi-mu}) are absolutely convergent for $|q|>1$.

\section{Evaluation of Braid Generators}
To begin we state
\vskip.in
\noindent {\bf Proposition 8.1} The universal $R$-matrix $R$ for
$U_q(\hat{\cal G})$ satisfies the conjugation relation: $R^\dagger=R^T$.
\vskip.in
\noindent {\bf Proof:} The proof is similar to the proof for quantum simple
Lie algebra case (see the Appendix in the first reference of \cite{Gould}),
thanks to the uniqueness of the universal $R$-matrix \cite{KT}.~~~$\Box$
\vskip.1in
An alternative proof of proposition 8.1 for
the nontwisted case, $U_q(\hat{\cal G})=U_q({\cal G}^{(1)})$, is given
in \cite{GZ} by using the
explicit expression for the universal $R$-matrix \cite{KT}\cite{ZG}.

Let $P$ be the permutation operator on $V(\Lambda)\otimes V(\Lambda)$ defined
by
$P(|\mu>\otimes |\nu>)=|\nu>\otimes |\mu>\,,~\forall |\mu>\,,\,|\nu>\in
V(\Lambda)$ and Let
\begin{equation}
\sigma=PR\,~~~~~\in\,{\rm End}(V(\Lambda)\otimes V(\Lambda))\label{pr}\;.
\end{equation}
Here and in what follows we regard elements of $U_q(\hat{\cal G})$ as
operators on $V(\Lambda)$. Then (\ref{hopf}) is equivalent to
\begin{equation}
\sigma\Delta(a)=\Delta(a)\sigma\,~~~~\forall a\in U_q(\hat{\cal G})\label{11}
\end{equation}
which means that $\sigma$ is an operator obeying the condition of proposition
2.1. Furthermore, with the help of proposition 8.1 one can easily show that
$\sigma$ is self-adjoint and thus may be diagonalized.

Recall that $\lim_{q\rightarrow 1}\;\sigma=P$ and $P$ is diagonalizable
on $V(\Lambda)\otimes V(\Lambda)$ with eigenvalues $\pm 1$.
Following \cite{Gould}, we define
the subspaces
\begin{equation}
W_\pm=\{x\in V(\Lambda)\otimes V(\Lambda)\,|\,
\lim_{q\rightarrow 1}\sigma x=\pm x\}\;.
\end{equation}
Since $\sigma$ is self-adjoint we may clearly write
\begin{equation}
V(\Lambda)\otimes V(\Lambda)=W_+\bigoplus W_-\;.
\end{equation}
Let $P[\pm]$ denote the projection operators defined by
\begin{equation}
P[\pm](V(\Lambda)\otimes V(\Lambda))=W_\pm\;.
\end{equation}
Since $\sigma$ is an $U_q(\hat{\cal G})$-invariant each subspace $W_\pm$
determines a $U_q(\hat{\cal G})$-module and $P[\pm]$ commute with the action
of $U_q(\hat{\cal G})$. We write
\begin{equation}
V(\Lambda)\otimes V(\Lambda)=
\bigoplus_{\hat{\nu}\in
D^+}\bar{V}(\hat{\nu})\label{decom1}
\end{equation}
where $\hat{\nu}\sim\Lambda+\nu\,,~~\nu\in\Pi(\Lambda)$ and
\begin{equation}
\bar{V}(\hat{\nu})=V(\hat{\nu})
\bigoplus\,\cdots\,\bigoplus
V(\hat{\nu})~~~~~~~(m_{\hat{\nu}}~ {\rm terms})
\end{equation}
with $m_{\hat{\nu}}$ being the multiplicity of the module $V(\hat{\nu})$ in
the tensor product decomposition (\ref{decom1}). Let now $P[\hat{\nu}]$ be
the projections:
\begin{equation}
P[\hat{\nu}] (V(\Lambda)\otimes V(\Lambda))=\bar{V}(\hat{\nu})\;.
\end{equation}
We then deduce the following decompositions:
\begin{equation}
W_\pm=\bigoplus_{\hat{\nu}\in D^+}\,\bar{V}_\pm(\hat{\nu})
\end{equation}
where
\begin{equation}
\bar{V}_\pm (\hat{\nu})=P[\pm]\bar{V}(\hat{\nu})
\end{equation}
Let
\begin{equation}
P[\hat{\nu};\pm]=P[\hat{\nu}]P[\pm]=P[\pm]P[\hat{\nu}]
\end{equation}
then
\begin{equation}
\bar{V}_\pm (\hat{\nu})=P[\hat{\nu};\pm] (V(\Lambda)\otimes
V(\Lambda))=P[\hat{\nu}]W_\pm\;.
\end{equation}

Observe
\begin{equation}
\sigma^2=PRP\cdot R=R^TR=(v\otimes v)\Delta(v^{-1})
\end{equation}
where we have used (\ref{vv1}) in the last step. It then follows from
(\ref{chi1}) that on the submodule $\bar{V}_\pm (\hat{\nu})$ the
$\sigma^2$ has the eigenvalue
\begin{equation}
\chi_{\hat{\nu}}(\sigma^2)=\chi_{\Lambda}(v)
\chi_{\Lambda}(v)\chi_{\hat{\nu}}(v^{-1})=
q^{(\hat{\nu},\hat{\nu}+2\rho)-2(\Lambda,\Lambda+2\rho)}\;.
\end{equation}
Since $\sigma$ is self-adjoint the above equation implies
\begin{equation}
\chi_{\hat{\nu}}(\sigma)=\pm
q^{(\hat{\nu},\hat{\nu}+2\rho)/2-(\Lambda,\Lambda+2\rho)}
\end{equation}
on $\bar{V}_\pm (\hat{\nu})$, respectively. We thus arrive at \cite{GZ} the
following spectral decomposition formula for $\sigma$ and its powers:
\begin{equation}
\sigma^l=q^{-l(\Lambda,\Lambda+2\rho)}\sum_{\hat{\nu}\in D^+}
 \, q^{l(\hat{\nu},\hat{\nu}+2\rho)/2}
 \cdot \left ( P[+]+(-1)^{l}P[-]\right ) P[\hat{\nu}],~~~~
 l\in {\bf Z}\label{sigma1}\;.
\end{equation}

For the special case discussed in this section that $R$ and thus
$\sigma$ are operators acting on tensored space of the same
representation, i.e. $R\,,~\sigma\in\, {\rm End}(V(\Lambda)\otimes
V(\Lambda))$,
Casimir invariants could be more easily constructed with
the help of the spectral decomposition formula for $\sigma$.
We have
\vskip.1in
\noindent{\bf Proposition 8.2:}
\begin{equation}
\bar{C}^\Lambda_l=(I\otimes{\rm tr})\{[I\otimes
\pi_\Lambda(q^{2h_\rho})]\sigma^l\}
\,,~~~~~l\in {\bf Z}^+
\end{equation}
are Casimir invariants acting on $V(\Lambda)$. Their eigenvalues
are given by
\begin{equation}
\chi_\Lambda(\bar{C}^\Lambda_l)=q^{-\frac{l}{2}(\Lambda,\Lambda+2\rho)}
  \sum_{\nu\in\Pi(\Lambda)}
 (m^+_{\nu}+(-1)^lm^-_{\nu})q^{\frac{l}{2}(\nu,\nu+2(\Lambda+\rho))}
 \frac{D_q[\Lambda+\nu]}{D_q[\Lambda]}
 \,,~~~~l\in {\bf Z}^+\label{chi-lambda}
\end{equation}
where $m^\pm_{\nu}\in
{\bf Z}^+$ are the mutliplicities of $V(\nu)$ in $W_\pm$,
respectively, so that $m_{\nu}=m^+_{\nu}+m^-_{\nu}$.
\vskip.1in
\noindent{\bf Proof:} This can be proven similarly to proposition 6.1 in
section 6 with the help of (\ref{11}) and the spectral decomposition
(\ref{sigma1}).~~~$\Box$
\vskip.1in
\noindent{\bf Remarks:} (i) The eigenvalue formula (\ref{chi-lambda}) is
completely determined by the weight spectrum of the irrep. $\pi_\Lambda$,
in contrast to the relevant formula in \cite{GZ}.\\
(ii) We emphasize that since $\sigma$ is an operator on
$V(\Lambda)\otimes V(\Lambda)$, the $\bar{C}^\Lambda_l$ are Casimir
operators acting {\em only} on $V(\Lambda)$. On the contrary, the $C^\Lambda_m$
in section 6 are Casimir operators which may act on an {\em arbitrary}
$V(\mu)$, regardless of $\mu=\Lambda$. However, when $\mu=\Lambda$ in
proposition 6.1, we have
$C^\Lambda_m=\bar{C}^\Lambda_{2m}$ since $\sigma^2=R^TR$. Therefore,
the Casimir $C^\Lambda_m$ are more general than $\bar{C}^\Lambda_l$.
(iii) The eigenvalue formuale (\ref{chi-lambda})
are absolutely convergent for $|q|>1$ (see Appendix).
\vskip.1in
Some straightforward works show that (\ref{chi-lambda}) can be rewritten
as the form
\begin{eqnarray}
\chi_\Lambda(\bar{C}^\Lambda_l)&=&q^{-\frac{l}{2}(\Lambda,\Lambda+2\rho)}
  \sum_{\nu\in M(\Lambda)}
  q^{\frac{l}{2}(\nu,\nu+2(\Lambda+\rho))}
  \frac{D_q[\Lambda+\nu]}{D_q[\Lambda]}\nonumber\\
& &\cdot\sum_{s=0}^\infty (m_{\nu-s\delta}^++(-1)^lm^-_{\nu-s\delta})\,
  q^{-s\{2g+l(g+2\kappa_\Lambda)\}}\,,~~~~l\in {\bf Z}^+\label{chi-lambda-s}
\end{eqnarray}
On the other hand, for $l=1$ we can prove by direct computation \cite{GZ}
\begin{equation}
\bar{C}^\Lambda_1=(I\otimes {\rm tr})\{[I\otimes\pi_\Lambda(q^{2h_\rho})]
  \sigma\}=q^{(\Lambda,\Lambda+2\rho)}\cdot I
\end{equation}
By comparing it with (\ref{chi-lambda-s}), we obtain the
interesting identity
\begin{equation}
q^{\frac{3}{2}(\Lambda,\Lambda+2\rho)}=\sum_{\nu\in M(\Lambda)}
 q^{\frac{1}{2}(\nu,\nu+2(\Lambda+\rho))}
 \frac{D_q[\Lambda+\nu]}{D_q[\Lambda]}
 \sum_{s=0}^\infty (m_{\nu-s\delta}^+-m^-_{\nu-s\delta})\,
  q^{-s(3g+2\kappa_\Lambda)}\label{id}\;.
\end{equation}

\section{Concluding Remarks}
We have obtained the Casimir invariants for the infinite-dimensional quantum
affine Lie algebra $U_q(\hat{\cal G})$ and computed their eigenvalues for
any integrable IHW representation. The eigenvalues are explicitly determined
by only a knowledge of the weight spectrum of the reference irrep and
are absolutely convergent for $|q|\,>\,1$.

In principal, the spectral decomposition formulae
(\ref{sigma1}) may be applied to define link polynomials explicitly
for a general integrable IHW representation of $U_q(\hat{\cal G})$.
However care must be taken in order to avoid
divergence issues since the inverse of the braid generator is unbounded
above. It would be of interest to find a way to overcome them.

Finally we remark that we may equivalently work with the coproduct and antipode
\begin{eqnarray}
&&\bar{\Delta}(q^h)=q^h\otimes q^h\,,~~~h=h_i,~d\,,~~~i=0,1,\cdots,
r\nonumber\\
&&\bar{\Delta}(e_i)=q^{h_i/2}\otimes e_i+e_i\otimes q^{-h_i/2}\nonumber\\
&&\bar{\Delta}(f_i)=
q^{h_i/2}\otimes f_i+f_i\otimes q^{-h_i/2}\nonumber\\
&&\bar{S}(a)=-q^{-h_\rho}aq^{h_\rho}\,,~~~a=e_i,f_i,h_i,d\label{coproduct2}
\end{eqnarray}
which are obtained by making the interchange $q\leftrightarrow q^{-1}$ in
(\ref{coproduct1}).
Then the universal R-matrix $R$ implements the change $R\leftrightarrow R^T$.
Carrying on the
similar calculations above, we are able to obtain another set of family of
Casimir invariants which are given by the similar formulae above with
$q\leftrightarrow q^{-1}$ and thus their eigenvalues
are absolutely convergent for $|q|\,<\,1$.

\section*{Appendix}
In this Appendix we prove that the eigenvalue formula (\ref{chi-mu})
is absolutely convergent for $|q|>1$. The same arguments are valid for
the absolute convergence of formula (\ref{chi-lambda}).

Let us first of all look at $\beta_\nu(\mu)$ in (\ref{chi-mu}),
\begin{eqnarray}
\beta_\nu(\mu)&=&(\Lambda,\Lambda+2\rho)-(\nu,\nu+2(\mu+\rho))\nonumber\\
&=&(\Lambda,\Lambda+2\rho)-(\nu,\nu+2\rho)+2(\Lambda-\nu,\mu)-2(\Lambda,\mu)
\end{eqnarray}
It follows that
\begin{equation}
\beta_\nu(\mu)+2(\Lambda,\mu)=(\Lambda,\Lambda+2\rho)-(\nu,\nu+2\rho)
+2(\Lambda-\nu,\mu)\geq 0
\end{equation}
Therefore, for $|q|>1$ the eigenvalue formula (\ref{chi-mu}) is dominated
by
\begin{equation}
q^{2m(\Lambda,\mu)}\sum_{\nu\in\Pi(\Lambda)}m_\nu\,
\cdot \frac{D_q[\mu+\nu]}{D_q[\mu]}\label{dominated}
\end{equation}
Now the sum in the above equation is easily seen to be equal to
$\chi_\mu(C^\Lambda_0)$
which according to the definition is nothing but
the $q$-dimension of the irrep $\pi_\Lambda$. Hence (\ref{dominated})
is absolutely convergent for $|q|>1$, thanks to the absolute convergence of
the $q$-dimension.~~~$\Box$
\vskip.1in
\noindent {\bf Remarks:} An alternative proof of the above result, which is
equivalent to the arguments used in \cite{GZ}, follows
from (\ref{chi-mu}) according to which we have
\begin{equation}
\chi_\mu(C^\Lambda_m)=\sum_{\stackrel{\nu\in\Pi(\Lambda)}{\mu+\nu\in D^+}}
m_{\mu+\nu}\,q^{-m\beta_\nu(\mu)}\frac{D_q[\mu+\nu]}
{D_q[\mu]}\label{appendix2}
\end{equation}
Now let
\begin{equation}
M_\mu(\Lambda)=\{\nu\in M(\Lambda)\,|\,\mu+\nu\in D^+\}
\end{equation}
which is a {\em finite} set. Then we may write (\ref{appendix2}) as
\begin{equation}
\chi_\mu(C_m^\Lambda)=\sum_{\nu\in M_\mu(\Lambda)}\sum_{s=0}^\infty
m_{\mu+\nu-s\delta}\,q^{-m\beta_{\nu-s\delta}(\mu)}
\frac{D_q[\mu+\nu-s\delta]}{D_q[\mu]}\label{appendix3}
\end{equation}
which, using (\ref{proof-d}) and (\ref{proof}), becomes
\begin{equation}
\chi_\mu(C_m^\Lambda)=\sum_{\nu\in M_\mu(\Lambda)}
q^{-m\beta_{\nu}(\mu)}
\frac{D_q[\mu+\nu]}{D_q[\mu]}
\sum_{s=0}^\infty
m_{\mu+\nu-s\delta}
q^{-2s\{g+m(g+\kappa_\Lambda+\kappa_\mu)\}}\label{appendix4}
\end{equation}
Recalling that $M_\mu(\Lambda)$ is a finite set and the multiplicities
\begin{equation}
m_{\mu+\nu-s\delta}=\sum_{\stackrel{\theta\in\Pi(\Lambda)}{\mu+\theta\sim
\mu+\nu-s\delta}}m_\theta\;{\rm sign}w_\theta
\end{equation}
where $w_\theta(\mu+\theta+\rho)=\mu+\nu+\rho-s\delta$, are finite integers,
we immediately see that the r.h.s. of (\ref{appendix4}) is absolutely
convergent for $|q|>1$.

\end{document}